\documentclass[floatfix,amssymb,prl,twocolumn,superscriptaddress,nofootinbib]{revtex4-1}
		
\usepackage{amssymb,amsmath,verbatim,mathtools,needspace,enumitem,etoolbox,graphicx,physics,microtype,afterpage,bm,soul}
\usepackage[dvipsnames, usenames]{xcolor}
\definecolor{linkcolor}{rgb}{0.0,0.3,0.5}
\usepackage[unicode, colorlinks=true, linkcolor=linkcolor, citecolor=linkcolor, filecolor=linkcolor,urlcolor=linkcolor, pdfusetitle]{hyperref}
\usepackage[all]{hypcap}
\usepackage[T1]{fontenc}
\usepackage[utf8]{inputenc}
\usepackage{tabularx}
\usepackage{float}
\usepackage{lmodern}
\allowdisplaybreaks
\usepackage{tikz}
\usepackage{color}
\usepackage{framed}
\usepackage{hyperref}
\hypersetup{colorlinks, citecolor=bluscuro, linkcolor=black, urlcolor=bluscuro}
\definecolor{rossos}{cmyk}{0,1,1,0.55}
\definecolor{bluscuro}{rgb}{0.15, 0.2, .85}
\definecolor{bluchiaro}{cmyk}{1,.3,0.,0.1}
\definecolor{ForestGreen}{rgb}{0.13, 0.55, 0.13}

\newcommand{\be}{\begin{equation}}
\newcommand{\ee}{\end{equation}}

\def\lsim{\mathrel{\rlap{\lower4pt\hbox{\hskip0.5pt$\sim$}}
    \raise1pt\hbox{$<$}}}         %less than or approx. symbol
\def\gsim{\mathrel{\rlap{\lower4pt\hbox{\hskip0.5pt$\sim$}}
    \raise1pt\hbox{$>$}}}         %greater than or approx. symbol

\begin{document}

\title{Gravitational memory and soft theorems: The local perspective}

\author{Valerio De Luca}
\email{vdeluca@sas.upenn.edu}
\affiliation{Center for Particle Cosmology, Department of Physics and Astronomy,
University of Pennsylvania 209 South 33rd Street, Philadelphia, Pennsylvania 19104, USA}

\author{Justin Khoury}
\email{jkhoury@sas.upenn.edu}
\affiliation{Center for Particle Cosmology, Department of Physics and Astronomy,
University of Pennsylvania 209 South 33rd Street, Philadelphia, Pennsylvania 19104, USA}

\author{Sam S. C. Wong}
\email{samwong@cityu.edu.hk}
\affiliation{Department of Physics, City University of Hong Kong, \\
Tat Chee Avenue, Kowloon, Hong Kong SAR, China}

% \date{\today}

\begin{abstract}
\noindent
In general relativity, gravitational memory describes the lasting change in the separation and relative velocity of freely falling detectors after the passage of gravitational waves (GWs). 
In this paper, we elucidate the relation between Bondi-Metzner-Sachs  transformations at future null infinity and the description of gravitational memory in local synchronous coordinates, commonly used
in GW detectors like LISA. We show that gravitational memory corresponds to large residual diffeomorphisms in this gauge, such as
volume-preserving spatial rescalings. We reproduce the associated soft theorems for scattering amplitudes. 
Finally, we derive novel soft theorems for equal-time (in-in) correlation functions, which are recognized as the flat space analogues of inflationary consistency relations with a soft tensor mode. These relations provide a pathway toward uncovering deeper connections between gravitational memory and cosmological correlators.

\end{abstract}

\maketitle

Gravitational memory effects describe a lasting change in the GW strain associated to transient GW sources. 
First predicted by Zel'dovich and Polnarev in linearized gravity~\cite{Zeldovich:1974gvh}, and further developed by Braginsky and Grishchuk~\cite{Braginsky:1985vlg},
memory effects were generalized to nonlinear general relativity (GR) in the early 90's~\cite{Christodoulou:1991cr, Blanchet:1992br,Thorne:1992sdb}. (See~\cite{PhysRevD.44.R2945, Favata:2008yd, Blanchet:2020ngx, Blanchet:2023pce, Cunningham:2024dog} for post-Newtonian/post-Minkowskian computations, and~\cite{Pollney:2010hs, Grant:2023jhd, Yoo:2023spi} for numerical simulations.) Both effects are commonly referred to as displacement memory,  affecting the separation and relative velocity between observers who are comoving before and after the GW burst. Later works revealed additional subleading memory effects, such as spin and center-of-mass memory~\cite{Nichols:2017rqr, Flanagan:2018yzh, Nichols:2018qac}, included in the broader class of persistent observables~\cite{Flanagan:2018yzh, Flanagan:2019ezo,Grant:2021hga}.

While current detectors like LIGO/Virgo/KAGRA may eventually detect memory effects only from very loud sources~\cite{Favata:2009ii, Johnson:2018xly, Hubner:2021amk}, future detectors like the Einstein Telescope~\cite{Punturo:2010zz,Branchesi:2023mws}, Cosmic Explorer~\cite{Reitze:2019iox}, DECIGO~\cite{Kawamura:2011zz, Kawamura:2020pcg}
and LISA~\cite{LISA:2017pwj, LISA:2022kgy} are likely to measure displacement memory both from individual events and population-based procedures~\cite{Favata:2010zu, Lasky:2016knh,Islo:2019qht, Hubner:2019sly, Boersma:2020gxx, Hubner:2021amk, Boersma:2020gxx, Grant:2022bla, Goncharov:2023woe, Gasparotto:2023fcg,Inchauspe:2024ibs, Hou:2024rgo}. Pulsar timing arrays will require more time for detection~\cite{Pshirkov:2009ak, NANOGrav:2019vto, Islo:2019qht, vanHaasteren:2022agf}. The detection of memory would offer yet another key experimental test of GR.

Gravitational memory is closely related to Bondi-Metzner-Sachs (BMS) asymptotic symmetries~\cite{1962RSPSA.269...21B, Sachs:1962wk, Sachs:1962zza, Strominger:2017zoo}.
Specifically, the permanent shift of the asymptotic shear after the passage of GWs can be equivalently characterized as a transition between two different asymptotic BMS frames related by a supertranslation~\cite{Strominger:2013jfa,Strominger:2014pwa,Ashtekar:2014zsa,Flanagan:2015pxa,Kehagias:2016zry}. Similarly, the extended BMS group is related to subleading spin and center-of-mass memory effects~\cite{Barnich:2009se, Barnich:2010eb, Kapec:2014opa, Nichols:2018qac, Strominger:2017zoo, Himwich:2019qmj}, while more extended versions are associated to additional memory-type effects~\cite{Campiglia:2015yka, Freidel:2021fxf, Godazgar:2018qpq, Compere:2018ylh, Seraj:2021rxd, Godazgar:2022pbx, Siddhant:2024nft}.

Both memory and asymptotic symmetries, together with Weinberg's soft graviton theorems~\cite{Weinberg:1965nx}, form an ``infrared triangle''~\cite{He:2014laa, Strominger:2014pwa, Strominger:2017zoo, Anupam:2018vyu, Himwich:2019qmj}. Each offers an equivalent description of gravity at large distances. The link between soft theorems and large residual diffeomorphisms (diffs) has been widely discussed in many contexts, including cosmology~\cite{Creminelli:2012ed, Kehagias:2013yd, Peloso:2013zw, Hinterbichler:2013dpa, Hui:2018cag, Hui:2022dnm}.

The precise relation between memory in the ``local'' coordinates of GW detectors, such as transverse-traceless (TT) gauge,
and BMS transformations, which act asymptotically, remains unclear. (See~\cite{Avery:2015gxa, Campiglia:2016efb} for related work in this direction.) The goal of this paper, and its more detailed companion~\cite{DeLuca:2024asq},
is to elucidate this relation. We determine the residual diffs in TT gauge that encode gravitational memory, and show their equivalence to BMS transformations when translated to TT gauge. Starting from the Ward identities for the local residual diffs, we then derive the associated soft theorems, for both scattering amplitudes and in-in correlators. A key result is that the latter are the flat space analogues of the familiar inflationary consistency relations with a soft tensor mode.

\vspace{0.1cm}
%----------------------------------------------------------------------------------------------------
\noindent{{\bf{\em Gravitational memory and BMS symmetry.}}}
%---------------------------------------------------------------------------------------------------- 
We begin by reviewing the relation between gravitational memory and BMS transformations.
Adopting Bondi coordinates, the metric takes the general form~\cite{Barnich:2009se, Barnich:2010eb, Strominger:2013jfa, Strominger:2017zoo,Grant:2021hga}
\begin{align}
{\rm d}s^2 &= - \left( 1 - \frac{2 V}{R} \right) e^{2\beta/R} {\rm d} U^2 - 2 e^{2\beta/R} {\rm d} U  {\rm d} R \nonumber \\
& + R^2 \mathcal{H}_{ab} \left( {\rm d} \theta^a - \frac{\mathcal{W}^a}{R^2} {\rm d} U \right) \left( {\rm d} \theta^b - \frac{\mathcal{W}^b}{R^2} {\rm d} U \right)\,,
\end{align}
%$(U,R, \theta^a)$
where~$U = T-R$ is retarded time, and~$\theta^a$ are angular coordinates. The metric functions~$\beta$,~$V$,~$\mathcal{W}_a$, and~$\mathcal{H}_{ab}$ are obtained by solving Einstein’s equations, subject to initial data on a null hypersurface and suitable boundary conditions at null infinity. (See, {\it e.g.},~\cite{Madler:2016xju}.)
Of particular interest is the metric~$\mathcal{H}_{ab}$ on the sphere, which in the large distance limit $R\to \infty$, keeping~$U$ fixed, becomes
\begin{equation}
\mathcal{H}_{ab}  = \gamma_{ab} + \frac{1}{R} C_{ab} (U,\theta^c) + {\cal O}\left(R^{-2}\right) \,,
\end{equation}
where~$\gamma_{ab}$ is the standard two-sphere metric, and~$C_{ab}$ is the shear tensor. 
The latter is intimately connected to the TT metric perturbation~$H_{ij}^\text{\tiny TT}$ via~\cite{Blanchet:2020ngx,Blanchet:2023pce}
\begin{equation}
\frac{G}{R} C_{ab} = e^i_{\langle a} e^j_{b\rangle} H_{ij}^\text{\tiny TT}\,. 
\label{CH}
\end{equation}
Here,~$e^i_a = \partial N^i/\partial \theta^a$ are two-sphere basis vectors,~$N_i = X_i/R$ is unit normal,
and~$e^i_{\langle a} e^j_{b\rangle} = e^i_{(a} e^j_{b)} - \frac{1}{2} \gamma_{ab}P^{ij}$, where~$P_{ij} = \delta_{ij} - N_i N_j$
is the projector onto the sphere. 

To see how the shear carries the information about gravitational radiation and memory, it is useful to work in the multipolar post-Minkowskian approximation~\cite{Blanchet:2020ngx}
and expand the TT metric perturbation as~\cite{Maggiore:2007ulw}
\begin{align}
\label{HijPN}
H_{ij}^\text{\tiny TT} & = \frac{4G}{R} \Pi_{ijkl}(N) \sum_{\ell=2}^\infty \frac{N_{L-2}}{\ell!} \Big\{ \mathcal{U}_{kl L-2} (U) \nonumber \\
& ~~~~~~~~~~ + N^s \epsilon_{ms(k} \mathcal{V}_{l)mL-2} (U) \Big\}\,,
\end{align}
where the radiative moments $\mathcal{U}_L,\mathcal{V}_L$ are symmetric, trace-free (STF) arrays, and~$\Pi_{ijkl} = \frac{1}{2} (P_{ik} P_{jl} + P_{il} P_{jk} - P_{ij}P_{kl})$ is the TT projector. We use the multi-index notation $L = i_1 i_2 \dots i_\ell$, with $\partial_L = \partial_{i_1} \dots \partial_{i_\ell}$ and $N_L = N_{i_1} \dots N_{i_\ell}$. 

The radiative multipole moments~$\mathcal{U}_L,\mathcal{V}_L$ carry information about the energy-momentum of the source responsible for the GW signal, and contain among its many contributions the memory terms~\cite{Thorne:1992sdb, Blanchet:1992br, Favata:2008yd}. 
Equation~\eqref{CH} shows that a change~$ \Delta H_{ij}^\text{\tiny TT}$ due to the memory implies a corresponding change~$\Delta C_{ab}$ in the shear. 
This relation enables us to understand how BMS symmetry transformations can be applied to describe memory effects~\cite{Strominger:2013jfa}.

BMS symmetries are diffs acting on future null infinity that preserve its intrinsic geometric properties~\cite{Wald:1984rg, Geroch:1977big}.
Their extension to the bulk is obtained by requiring that they maintain the retarded Bondi gauge conditions. The corresponding diff has the asymptotic form
\begin{align}
\label{BMSdiffgeneral}
\xi^U & = f(U, \theta^a) \equiv T(\theta^a) + \frac{U}{2}  D_aY^a(\theta^b)\,;  \nonumber \\ 
\xi^R & = -\frac{R}{2} D_aY^a + \frac{1}{2} D^2 f + \mathcal{O} \left(R^{-1} \right)\,;  \nonumber \\
\xi^a & = Y^a - \frac{1}{R} D^a f + \mathcal{O} \left(R^{-2} \right)\,,
\end{align}
where~$D_a$ is the covariant derivative on the two-sphere, and~$D^2$ is the corresponding Laplacian.
The scalar function~$T(\theta^a)$ appearing in the time diff generates supertranslations, while~$Y^a$'s are globally-defined conformal
Killing vectors on the sphere.\footnote{In extended versions of the BMS group, the generators~$Y^a$ include all local conformal transformations and diffs on the sphere~\cite{Barnich:2009se}.}

Under the diff~\eqref{BMSdiffgeneral}, the shear tensor transforms as~\cite{Blanchet:2020ngx,Blanchet:2023pce}
\begin{align}
 \delta C_{ab} & = - 2 D_aD_b f + \gamma_{ab} D^2 f \nonumber \\
 & = e^i_{\langle a} e^j_{b\rangle}  \sum_{\ell =2 }^{\infty} \ell (\ell-1) N_{L-2} f_{ij L-2}(U)\,,
\end{align}
where in the last step we have performed the STF decomposition:~$f(U, \theta^a) = \sum_{\ell} N_L f_L(U)$.
We can compare this result with the expansion of~$C_{ab}$ in terms of~${\cal U}_{L}(U)$ and~${\cal V}_{L}(U)$,
obtained by combining~\eqref{CH} and~\eqref{HijPN}. Since~$f(U, \theta^a)$ is linear in~$U$ according to Eq.~\eqref{BMSdiffgeneral},
we see that a BMS transformation allows us to remove the constant and linear-in-$U$ terms in the radiative multipole moments,
via the identification (for~$\ell \geq 2$)
\begin{equation}
     f_{ij L-2}(U) = \frac{4}{\ell (\ell-1) \ell!}  {\cal U}_{ij L-2}^\text{\tiny lin}(U) \,,
\end{equation}
where ``lin'' indicates terms at most linear in~$U$. Notice that the odd parity term $\cal V$ cannot be fixed, since the BMS function $f = T + \frac{U}{2}D_a Y^a$ is composed of a scalar and the divergence of a vector.

\vspace{0.1cm}
%----------------------------------------------------------------------------------------------------
\noindent{{\bf{\em Residual diffs in TT coordinates.}}}
%----------------------------------------------------------------------------------------------------
Having reviewed the connection between gravitational memory and BMS transformations in Bondi coordinates,
we now turn to the description in TT gauge, which is the familiar ``local'' coordinate system of GW detectors~\cite{Maggiore:2007ulw,DeLuca:2019ufz,Lee:2024oxo}.
We will determine the residual diffs that encode gravitational memory in this gauge, and show their equivalence to
BMS transformations when translated to TT coordinates.

Consider a GW source located far away from the detector. The large distance between source and detector allows us to focus on the leading~$1/R$ contribution to the GW strain. Adopting harmonic coordinates around the detector, the Bondi coordinates can be expanded in its vicinity as~$X^i = \bar{X}^i + x^i$,~$U = \bar{U} + u$, where bars indicate the detector center-of-mass location and average observation time, while small letters indicate perturbations around them, with~$u = t - \bar{N}_k x^k$. Without loss of generality, we henceforth set~$\bar{U} = 0$.

To leading order in~$1/\bar{R}$, the GW strain~\eqref{HijPN} describes a plane wave propagating in the~$\bar{N}^i$ direction:
\begin{align}
H_{ij}^\text{\tiny TT} & =   \frac{4G\overline{\Pi}_{ij nr}}{\bar{R}}  \sum_{\ell, k }^\infty \frac{\bar{N}_{L-2}}{\ell!\,k!} \left[ {\cal U}^{(k)}_{nr L-2} + \bar{N}^s\epsilon_{ms(n}  {\cal V}^{(k)}_{r)mL-2} \right] u^k  \nonumber \\
& \equiv \frac{1}{\bar{R}} \Big[ A_{ij}(\bar{N}) + B_{ij}(\bar{N}) u + \ldots\Big]\,,
\label{hij TT local}
\end{align}
where~${\cal U}^{(k)} \equiv \frac{\partial^k}{\partial U^k} {\cal U}(U) \big\vert_{U=0}$. 
The leading-order tensors~$A_{ij}$ and~$B_{ij}$, which are both traceless and transverse to~$\bar{N}$,
encode the displacement memory effect. As expected from the equivalence principle, both terms can be removed
with a suitable gauge-preserving diff:
\begin{equation}
\xi_i = - \frac{1}{2\bar{R}}  \Big(A_{ij}  + B_{ij} u \Big) x^j - \frac{1}{4\bar{R}} B_{jk} \bar{N}_i x^j x^k\,.
\label{TT space diff}
\end{equation}
Note that~$\xi_i$ satisfies~$\partial^i\xi_i = 0$ and~$\vec{\nabla}^2 \xi_i = 0$, which ensures that~$\delta H_{ij}^\text{\tiny TT} = \partial_i \xi_j + \partial_j \xi_i$ remains TT. The terms linear in~$x^j$ describe a time-dependent anisotropic rescaling, while the quadratic term in~$\vec{x}$ removes
a homogeneous acceleration. Because~$\xi^i$ is time-dependent, a compensating, spatially-dependent time translation,
\begin{equation}
\xi_0 = \frac{1}{4\bar{R}} B_{ij} x^i x^j \,,
\label{TT time diff}
\end{equation}
is required to preserve the TT condition~$H_{0i}^{\rm TT}  = 0$.

Our next task is to establish the equivalence of the local residual diffs~\eqref{TT space diff}-\eqref{TT time diff}
with BMS transformations. For starters, it is useful to express the generalized BMS diff~\eqref{BMSdiffgeneral} in
Minkowski coordinates:
\begin{align}
\xi_0^\text{\tiny BMS} &= \frac{R}{2} D_a Y^a - \frac{1}{2} \left(D^2 + 2\right) f\,;   \\
\nonumber
\xi_i^\text{\tiny BMS} &=   \left( -\frac{R}{2} D_a Y^a + \frac{1}{2} D^2 f \right) N^i + e^i_a \Big(RY^a -D^a f\Big) \,,
\label{Mink diff}
\end{align}
Furthermore, the generator~$Y_a$ admits a Helmhotz-Hodge decomposition in terms of two scalars $\phi$ and $\psi$ as
\be
Y_a = D_a \phi - \varepsilon_a^{\;c}D_c\psi  = e^i_a R \Big(\partial_i \phi+ \epsilon_{ijk} N^j \partial_k \psi\Big)\,,
\ee 
where~$\varepsilon_{ac}$ is the Levi-Civita tensor on the sphere, such that~$f = T - \frac{U}{2} D^2 \phi$. This diff generates the long mode~$\delta h^\text{\tiny BMS}_{\mu \nu} = \partial_\mu \xi_\nu^\text{\tiny BMS} + \partial_\nu \xi_\mu^\text{\tiny BMS}$, whose full expression is given in~\cite{DeLuca:2024asq}. In the vicinity of the detector, it can be expanded up to
linear order as
\begin{align}
\delta h^\text{\tiny BMS}_{\mu\nu} & = \hat{H}_{\mu\nu} + \bar{P}_{km} H_{\mu\nu m} x^k + Q_{\mu\nu} u  + \ldots 
\label{h0i expand}
\end{align}
In general, this takes us out of TT gauge, and one must perform a compensating diff to restore TT gauge. For the purpose of this short paper,
we outline the explicit calculation for the constant piece~$\hat{H}_{\mu\nu}$, leaving the discussion of the linear-gradient terms to the
long paper~\cite{DeLuca:2024asq}. Expanding everything in terms of STF tensors, we find that~$\hat{H}_{\mu\nu}$ has components
\begin{align}
\nonumber
& \hat{H}_{0i} = \frac{1}{2} \sum_\ell \ell(\ell -1)(\ell +2 ) \\
& ~~~~~~~~ \times \bigg( \bar{N}_i  (\ell + 1)  \bar{N}_L \phi_L + \frac{1}{\bar{R} } \bar{P} _{in} \bar{N}_{L-1} T_{nL-1} \bigg)\,; \nonumber \\
\nonumber
& \hat{H}_{ij} =  \sum_\ell \ell(\ell -1)\Bigg\{-  \frac{1}{2} \bar{N}_i \bar{N}_j (\ell + 1) (\ell+2) \bar{N}_L \phi_L \\
\nonumber 
& -   \frac{1}{2\bar{R}} \Big(\bar{P}_{in} \bar{N}_j  + \bar{P}_{jn} \bar{N}_i \Big)  (\ell +2) \bar{N}_{L-1}  T_{nL-1}  \\
& + 2  \overline{\Pi}_{ijnr} \bar{N}_{L-2} \left[\phi_{nrL-2} - \frac{ T_{nrL-2}}{\bar{R}} + N^s  \epsilon_{ms(n} \psi_{r)m L-2} \right] \Bigg\} \,.
\label{Hij}
\end{align}
Note that~$\hat{H}_{00}$ is time independent; hence, it can be absorbed into a redefinition
of the Newtonian potential. 

\begin{figure}
    \centering
    \includegraphics[width=1\linewidth]{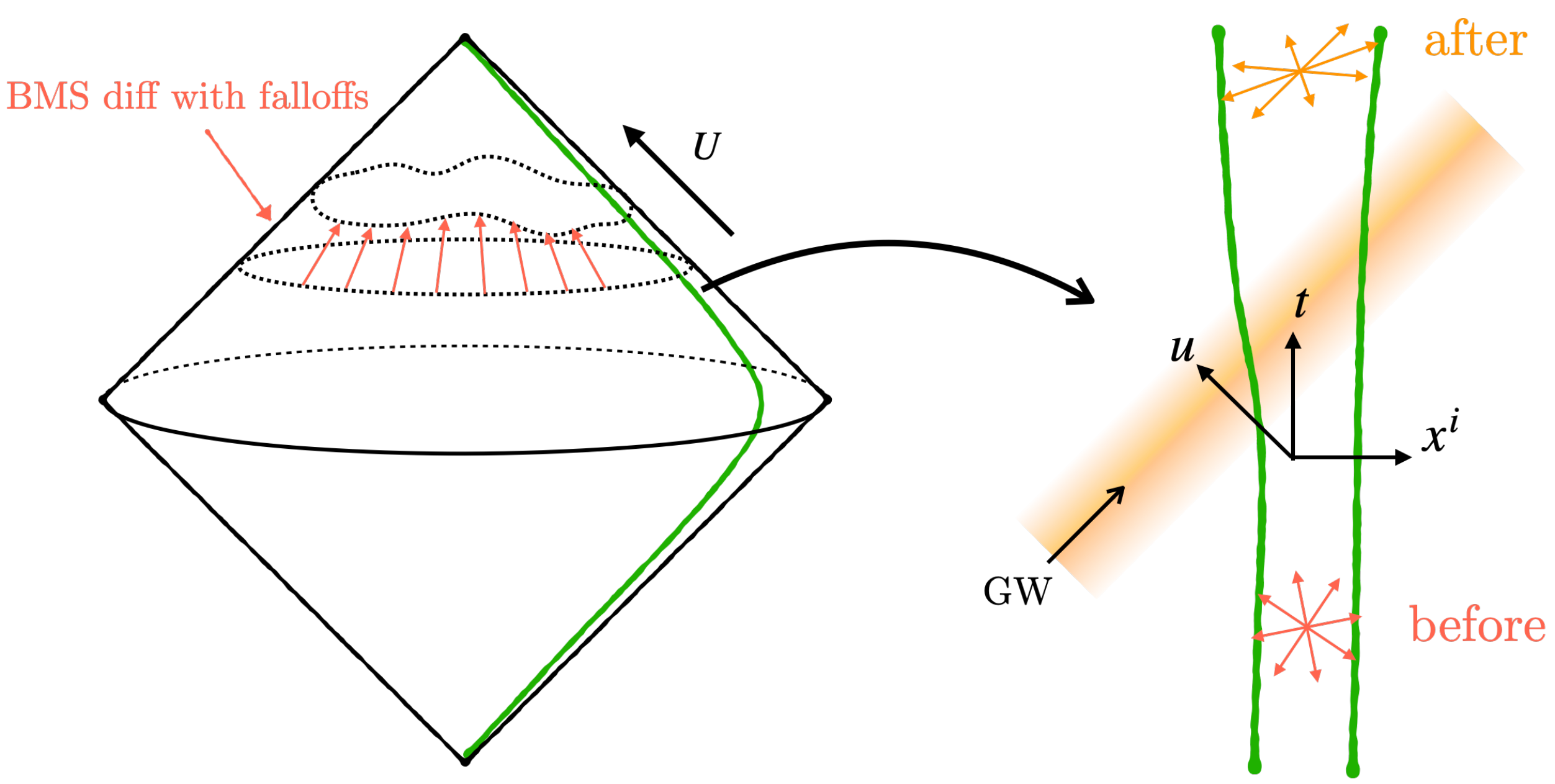}
    \caption{Illustrative representation of the equivalence between the BMS diff and the coordinate transformation associated to the detector in the TT gauge. The green lines denote the geodesics of two freely falling detectors measuring a GW memory effect. The BMS diff is equivalent to a combination of various coordinate transformations, denoted schematically by red arrows.}
    \label{fig:BMStoTT}
\end{figure}

To restore TT gauge, we must perform a compensating diff such that~$\hat{H}_{0i}$ is set to zero,
while~$\hat{H}_{ij}$ becomes traceless and transverse to~$\bar{N}$. To make~$\hat{H}_{0i}$ vanish
requires a {\it time-dependent spatial translation}, given by~$ \xi_i^\text{\tiny tran} = -\hat{H}_{0i} t$.
Next, notice that only the last line of Eq.~\eqref{Hij} is traceless and transverse to~$\bar{N}$.
The remainder can be removed by a suitable {\it spatial rescaling},~$\xi_i^\text{\tiny scaling} = \frac{1}{2} \left(\overline{\Pi}_{ijnr}\hat{H}_{nr} -\hat{H}_{ij}\right) x^j$. This leaves us with the last line of Eq.~\eqref{Hij}, which matches the constant term in  Eq.~\eqref{hij TT local}
with the identification
\begin{align}
\phi_L - \frac{T_{L}}{\bar{R}}   = \frac {2G}{\bar{R}\,\ell(\ell-1) \ell !}\, {\cal U}^{(0)}_{L}\,;~~ 
\psi_{L} =  \frac {2G}{\bar{R}\,\ell(\ell-1) \ell !}\, {\cal V}^{(0)}_{L}  \,. 
\label{constant identification}
\end{align}
To summarize, a constant TT mode~$\hat{H}_{ij}^\text{\tiny TT}$, which is the leading term in a coordinate expansion in the vicinity of the detector,
can be thought of as being induced by a BMS diff together with a compensating time-dependent translation and spatial rescaling,
\be
\xi_i = \xi_i^\text{\tiny BMS} + \xi_i^\text{\tiny tran} + \xi_i^\text{\tiny scaling} \,.
\ee
This is completely equivalent to leading residual diff in TT gauge, given by the~$A_{ij}$ piece in Eq.~\eqref{TT space diff}. Similarly, as shown in detail in~\cite{DeLuca:2024asq}, the linear-gradient TT mode can be generated by a BMS transformation and suitable compensating diff to restore TT gauge. The result matches the subleading residual diffs in TT gauge, given by the~$B_{ij}$ terms in Eqs.~\eqref{TT space diff} and~\eqref{TT time diff}. The equivalence between BMS and residual TT diffs is illustrated in Fig.~\ref{fig:BMStoTT}.

\vspace{0.1cm}
%----------------------------------------------------------------------------------------------------
\noindent{{\bf{\em Soft theorems for scattering amplitudes.}}}
%----------------------------------------------------------------------------------------------------
The explicit form of the large residual diffs in TT gauge, given by Eqs.~\eqref{TT space diff} and~\eqref{TT time diff}, and their relation to the asymptotic BMS symmetries, allows us to explore the third corner of the infrared triangle, {\it i.e.}, soft theorems. As noted earlier, the residual TT diffs share the same origin as the large diffs discussed in cosmology, which give rise to soft theorems for cosmological correlators~\cite{Creminelli:2012ed,Kehagias:2013yd,Peloso:2013zw, Hinterbichler:2013dpa, Hui:2018cag, Hui:2022dnm, Strominger:2013jfa, Creminelli:2024qpu}. In the following, we will derive the Ward-Takahashi identities associated with the local diffs around the detector and show
that they constrain the soft limits of scattering amplitudes with a soft graviton mode, leaving the discussion of equal-time correlators to the next section. 

For concreteness, we focus on the case where the hard-momentum modes are generic matter fields, denoted by~$\Phi$. To derive the identities, let us introduce the current $Q^{\mu}=\xi_{\alpha} T^{\alpha \mu}$, where~$\xi_{\alpha}$ is given by Eqs.~\eqref{TT space diff}-\eqref{TT time diff}, and $T^{\alpha \mu}$ includes the matter and pseudo energy-momentum tensor. It follows that
\begin{align}
   & {\rm i}\left[Q^0(t, \vec{x}), \Phi(t, \vec{y}) \right] = \delta^{(3)}(\vec{x}-\vec{y}) \xi^{\mu}\partial_{\mu} \Phi \,,
\end{align}
where we recognize the Lie derivative~$\mathcal{L}_\xi \Phi =  \xi^{\mu}\partial_{\mu}\Phi = \delta \Phi$ on the right-hand side.
The associated Ward identity is readily obtained~\cite{Weinberg:1995mt,DiVecchia:2015jaq}
\begin{align}
\label{eqdcomm}
&    {\rm i} \partial_{\mu} \left\langle Q^{\mu}(x)  \Phi(x_1)\dots \Phi(x_N)\right\rangle - {\rm i} \left\langle \partial_{\mu} Q^{\mu}(x)  \Phi(x_1)\dots \Phi(x_N)\right\rangle  \nonumber \\
   &= \sum_{m=1}^N \delta^{(4)}(x-x_j) \left\langle\Phi(x_1)\dots \delta\Phi(x_m)\dots  \Phi(x_N)\right\rangle\,
\end{align}
where~$\left\langle\cdots\right\rangle$ denotes time-ordered correlators. Such an equation can be Fourier transformed, introducing the momentum $q^2 = 0$ associated to the soft mode.

The first term on the left-hand side can be disregarded, if we retain only terms up to order $\mathcal{O}(q)$ and note that the correlator lacks a pole at $q=0$~\cite{DiVecchia:2015jaq}. To simplify the second term, we apply the Schwinger-Dyson equation associated with Einstein's equations in TT gauge,~$-2 \kappa T_{\mu \nu}  = \Box h_{\mu \nu}$, where~$\kappa^2 = 8 \pi G$. Its Fourier transform reduces to a Lehmann–Symanzik–Zimmermann (LSZ) operation~\cite{1955NCimS...1..205L} to obtain an external graviton state. The last step consists of performing a LSZ reduction on both sides of Eq.~\eqref{eqdcomm}, which allows us to write down the Ward identities up to $\mathcal{O}(q)$. The detailed derivation can be found in Ref.~\cite{DeLuca:2024asq}.

The diffs in Eqs.~\eqref{TT space diff} and \eqref{TT time diff} allow us to remove the constant~($A_{ij}$) and linear-gradient~($B_{ij}$) of the tensor mode. The associated Ward identities give rise to leading and subleading soft theorems. Denoting by~${\cal T}^{\mu\nu}(q;k_1, \dots, k_N )$ the on-shell only out-amplitude with a soft graviton mode~$h^{\mu\nu}(q)$, and by~${\cal T}(k_1, \dots, k_N)$ the off-shell amplitude~\cite{DiVecchia:2015jaq} without the soft graviton, the soft theorems are given by:
\begin{widetext}
\begin{align} \label{eqn:idenn=12}
\nonumber
& {\rm leading:} \quad \frac{\kappa^{-1}}{2}A^{ij} {\cal T}_{ij}(q; k_1, \dots, \bar{k}_N)  =  \sum_{m=1}^N A^{ij} \bigg[  \frac{ (k_m+q)_i(k_m+q)_j}{k_m\cdot q} +  \frac{ k_{m \,i} k_{m \, j}}{k_m\cdot q}  q^{\mu}  \frac{\partial}{\partial k_{m}^{\mu} } - k_{m \, i} \frac{\partial }{\partial k_{m}^j} \bigg] {\cal T}(k_1, \dots, \bar{k}_N)\,; \\
& {\rm subleading:} \quad \kappa^{-1}  M^{(\mu\nu) \alpha}  \frac{\partial}{\partial q_{\alpha}} {\cal T}_{\mu\nu}(q;k_1, \dots, \bar{k}_N) = - \sum_{m=1}^N  M^{\mu \alpha\beta } (k_m+q)_\mu  \left[ 2 \frac{(k_m+q)_{\alpha}(k_m+q)_{\beta}}{( k_m \cdot q)^2}\left( 1 + q^{\nu} \frac{\partial}{\partial k_{m}^\nu} \right) \right. \nonumber \\
& \hspace{8cm} \left. - 2\frac{(k_m + q)_\alpha}{k_m\cdot q} \frac{\partial}{\partial k_{m}^\beta} + \frac{\partial^2 }{\partial k_{m}^\alpha \partial k_{m}^\beta }  \right] {\cal T}(k_1, \dots, \bar{k}_N)\,,
\end{align}
\end{widetext}
where the array~$M_{\mu \alpha\beta}$ in the subleading identity has components~$M_{ijk} = B_{ij} \bar{N}_k + B_{ik} \bar{N}_j - B_{jk} \bar{N}_i$,~$M_{ij0} = M_{i0j} = M_{0ij} = - B_{ij}$. In these relations we have used momentum conservation on the last leg to set $\bar{k}_N = -\sum\limits_{l=1}^{N-1} k_l-q$. Importantly, the leading pole in $q$, $\frac{1}{k_m\cdot q}$, reproduces exactly Weinberg's soft factor for emitting/absorbing a soft graviton~\cite{Weinberg:1995mt}. 
 It is worth commenting that the result~\eqref{eqn:idenn=12} is similar to that obtained in Ref.~\cite{Hamada:2018vrw}, which share the same symmetries in the graviton sector in TT gauge. Our result displays the identity with the soft factor after removing delta functions and going  explicitly through the LSZ procedure~\cite{1955NCimS...1..205L}.

Finally, let us stress that our discussion on subleading soft theorems is restricted to tree level~\cite{Lysov:2014csa,Campiglia:2014yka, Campiglia:2016hvg}. As shown in Refs.~\cite{Laddha:2018myi, Sahoo:2018lxl, Laddha:2018vbn,  Campiglia:2019wxe, Saha:2019tub, Krishna:2023fxg}, loop order corrections induce a relevant logarithmic term in the subleading soft graviton theorem, which is associated to a different fall-off behavior of the shear tensor (proportional to $1/U$) and to further gravitational memory effects, known as tail-of-memories~\cite{Laddha:2018vbn,Ghosh:2021bam, Geiller:2024ryw, Trestini:2022tot, Trestini:2023ssa,Donnay:2022hkf, Agrawal:2023zea, Choi:2024ygx, Choi:2024ajz}. 
In particular, the corresponding Ward identities allow to reproduce the classical logarithmic soft graviton theorem, upon considering a dressing of the massive hard modes induced by their long-range gravitational interactions with the soft graviton~\cite{Giddings:2019hjc}.
In our derivation, while we account for the effects of superrotations (responsible for the tree-level subleading soft theorem), we do not perform the  dressing of the energy-momentum tensor of the massive hard fields in the diff's current $Q^\mu$. Performing such extension to study the effects of loop corrections is left to future work.

\vspace{0.1cm}
%----------------------------------------------------------------------------------------------------
\noindent{{\bf{\em Soft theorems for cosmological correlators.}}}
%----------------------------------------------------------------------------------------------------
Consistency relations in cosmology, analogously to soft theorems for scattering amplitudes in high energy physics, are exact symmetry statements  relating a $(N + 1)$-point correlation point with an external soft mode to a $N$-point function without the soft mode, based on some nonlinearly realized symmetry. They hold for both scalar and tensor soft modes~\cite{Maldacena:2002vr,Creminelli:2004yq,Cheung:2007sv,Creminelli:2012ed, Creminelli:2024qpu}, and have been studied in various contexts, such as inflation~\cite{Maldacena:2002vr, Creminelli:2004yq, Cheung:2007sv} and large scale structure~\cite{Kehagias:2013yd,Peloso:2013zw}. Furthermore, they have also been derived for asymptotic symmetries of cosmological spacetimes~\cite{Mirbabayi:2016xvc, Ferreira:2016hee, Hinterbichler:2016pzn}. It is therefore natural to show the existence of these relations for the local diffs of Eqs.~\eqref{TT space diff}-\eqref{TT time diff}. 

For concreteness, we focus first on the leading-order contribution, describing an anisotropic spatial rescaling $\xi^i = M^{i}_{\,j} x^j$. The matrix~$M_{ij}$ is symmetric and traceless, and can remove the long-wavelength part of a gravitational wave~$h_{ij}$ in TT gauge,~$ h_{ij} \to \bar{h}_{ij} = 2M_{ij}$, which mimics a constant memory term. 

Consider a general equal-time operator~${\cal O}(x_1,\ldots,x_N)$ build out of~$N$ matter fields~$\Phi$. Focusing on the time-independent part of the long mode, the correlation of~${\cal O}$ with a soft mode should be equivalent to the correlation of~${\cal O}$ in the transformed coordinate~$\tilde{x}^{i} =x^{i}+ \frac{1}{2} \bar{h}^{i}_{\,j} x^j$:
\begin{align}
\label{OO}
\hspace{-0.3cm}\left\langle {\cal O}(x_1,\ldots,x_N)\right\rangle_{h \to {\rm const.}} = \left\langle{\cal O}(\tilde{x}_1,\ldots,\tilde{x}_N)\right\rangle\,,
\end{align}
where~$\langle \ldots \rangle_{\bar{h}}$ and~$\langle \ldots \rangle$ denote the correlators with and without the long mode, respectively.
Expanding the right-hand side up to linear order in~$\bar{h}_{ij}$, we obtain
\begin{align}
    \left\langle \mathcal{O}(\tilde{x}_1,\ldots, \tilde{x}_N)\right\rangle  =  \left(1 + \frac{\bar{h}^k_{\,\ell}}{2} \sum_{m = 1}^N x_m^\ell \frac{\partial}{\partial x_m^k} \right)\left\langle{\cal O}(x_1\ldots,x_N)\right\rangle\,. 
\end{align}
Correlating both sides of Eq.~\eqref{OO} with~$h_{ij}$ gives
\begin{align}
\label{CRcorrelators}
 & \lim_{h \to {\rm const.}} \left\langle h_{ij}{\cal O}(x_1,\ldots,x_N) \right\rangle \\ 
\nonumber 
 & =   \lim_{h \to {\rm const.}} \frac{1}{2} \left\langle h_{ij} \bar{h}_{k \ell}\right\rangle \sum_{m = 1}^N x_m^\ell \frac{\partial}{\partial x_m^k} \left\langle{\cal O}(x_1,\ldots,x_N)\right\rangle\,.
\end{align}
To ensure that~$h$ is correlated with the mode~$\bar{h}$ being removed, we have assumed in the last step
that the time dependence of~$h$ in the soft limit matches that of~$\bar{h}$. This is the physical mode condition~\cite{Hui:2018cag}.
Furthermore, in order that the shift in~$h_{ij}$ corresponds to the long-wavelength limit of a physical mode,~$\bar{h}_{ij}$ should
satisfy the transversality condition $\hat{q}^i \bar{h}_{ij} (\hat{q}) = 0$, analogously to the ``adiabatic'' mode condition in cosmology~\cite{Hinterbichler:2013dpa}. Using this, the Ward identity~\eqref{CRcorrelators} can be expressed in Fourier space
in terms of the complex operator~${\cal O}\big(\vec{k}_1,\dots, \vec{k}_N\big) = \Phi\big(\vec{k}_1\big)\cdots \Phi\big(\vec{k}_N\big)$:
\begin{align}
\nonumber
&    \lim_{\vec q \to 0} \Pi^{ijk\ell}(\hat{q}) \frac{1}{P_h(q)} \left\langle h_{k\ell}\big(\vec q\big)  \Phi\big(\vec{k}_1\big)\cdots \Phi\big(\vec{k}_N\big)\right\rangle_{\rm c}' \\
    & ~   = -  \Pi^{ijk}_{~~\,\; \ell}(\hat{q})\sum_{m = 1}^N k_m^\ell \frac{\partial}{\partial k_m^k} \left\langle \Phi\big(\vec{k}_1\big)\cdots \Phi\big(\vec{k}_N\big)\right\rangle_{\rm c}'\,,
\label{N=2 identity with scalars}
\end{align}
where the subscript ``c'' denotes the connected part of correlators, and primes indicate the removal of three-dimensional Dirac delta functions enforcing momentum conservation. Lastly,~$P_h(q)$ is the tensor power spectrum, defined as~$\left\langle h_{ij} \big(\vec{q}\big) h_{ij} \big(\vec{q}\,'\big)\right\rangle = (2\pi)^3 \delta^{(3)} \big(\vec{q} + \vec{q}\,'\big) 4 P_h(q)$.  Equation~\eqref{CRcorrelators} represents the consistency relations for correlators under the anisotropic spatial rescaling symmetry in TT coordinates.

Equation~\eqref{N=2 identity with scalars} can be easily recognized as the flat-space analogue of tensor soft theorems in inflationary cosmology. In particular, when a long-wavelength primordial graviton mode $h_{k \ell}$ is present during inflation, it induces a deformation of spatial distances which can be viewed as a residual gauge transformation acting non-trivially on the short-scale curvature perturbations $\zeta$, but leaving the background de Sitter solution invariant. As a result, in single-field slow-roll models of inflation, correlators of the curvature perturbation are constrained according to the relation~\cite{Maldacena:2002vr, Creminelli:2004yq, Hinterbichler:2013dpa}
\begin{align}
\nonumber
&    \lim_{\vec q \to 0} \Pi^{ijk\ell}(\hat{q}) \frac{1}{P_h(q)} \left\langle h_{k\ell}\big(\vec q\big)  \zeta\big(\vec{k}_1\big)\cdots \zeta\big(\vec{k}_N\big)\right\rangle_{\rm c}' \\
    & ~   = -  \Pi^{ijk}_{~~\,\; \ell}(\hat{q})\sum_{m = 1}^N k_m^\ell \frac{\partial}{\partial k_m^k} \left\langle \zeta\big(\vec{k}_1\big)\cdots \zeta\big(\vec{k}_N\big)\right\rangle_{\rm c}'\,,
\label{inflation}
\end{align}
valid to lowest order in tensors. 
This result stems from the nonlinearly realized symmetry associated with spatial diffeomorphisms in de Sitter space, particularly through the anisotropic scaling induced by the soft graviton. Given that these relations are valid for de Sitter, their trivial generalization to flat spacetime shows the direct connection among the inflationary soft theorems and those derived in the context of gravitational memory.

Similarly to the results for scattering amplitudes, we can generalize the soft theorems for cosmological correlations by including the contribution from the subleading local diff.
To do so, we will follow similar steps to the ones considered above, and extend the background mode procedure outlined in Eq.~\eqref{OO} to
\begin{align}
\label{OOsub}
\hspace{-0.3cm}\left\langle {\cal O}(x_1,\ldots,x_N)\right\rangle_{h \to A + B u} = \left\langle{\cal O}(\tilde{x}_1,\ldots,\tilde{x}_N)\right\rangle\,,
\end{align}
where we now wish to remove the linear gradient contribution of the long-wavelength part of a gravitational
wave $h_{ij}$ in TT gauge, $h_{ij} \to A_{ij} + B_{ij} u$ (see Eq.~\eqref{hij TT local}). Here, we have to include the subleading contribution of the diff to the transformed coordinates as $\tilde{x}^\mu = x^\mu + \xi^\mu = x^\mu + M^\mu_{\,\,\, \nu \rho} x^\nu x^\rho$. Then, the right-hand side of~\eqref{OOsub} reads
\begin{align}
    & \left\langle \mathcal{O}(\tilde{x}_1,\ldots, \tilde{x}_N)\right\rangle  \nonumber \\
    & = \left(1 + M^\mu_{\,\,\, \nu \rho} \sum_{m = 1}^N x_m^\nu x_m^\rho \frac{\partial}{\partial x_m^\mu} \right)\left\langle{\cal O}(x_1\ldots,x_N)\right\rangle\,, 
\end{align}
where we have taken into account the linear variation of the scalar fields building the equal-time operator $\mathcal{O}$.
Correlating both sides of Eq.~\eqref{OOsub} with~$h_{ij}$ gives
\begin{align}
\label{CRcorrelators-sub}
 & \lim_{h \to A + B u} \left\langle h_{ij}{\cal O}(x_1,\ldots,x_N) \right\rangle \\ 
\nonumber 
 & =   \lim_{h \to A + B u}  \left\langle h_{ij} M^\mu_{\,\,\, \nu \rho} \right\rangle \sum_{m = 1}^N x_m^\nu x_m^\rho \frac{\partial}{\partial x_m^\mu} \left\langle{\cal O}(x_1,\ldots,x_N)\right\rangle\,.
\end{align}
In analogy to the physical mode condition discussed below Eq.~\eqref{CRcorrelators}, even at the subleading order it is assumed that the time dependence of $h$ matches the one of $M x \sim B x$ (see below~\eqref{eqn:idenn=12}), to ensure that their correlation is nonvanishing. In other words, one has assumed that the gravitational wave $h_{ij}$ and the hard modes $\Phi$ are evaluated at the same final time $u$.
By expanding the explicit form of $M^\mu_{\,\,\, \nu \rho}$, we can rewrite~\eqref{CRcorrelators-sub} as
\begin{align}
\label{CRcorrelators-sub-v2}
& \frac{\partial}{\partial u}  \langle h_{ij} {\cal O}(x_1,\ldots,x_N) \rangle \nonumber \\
& = \frac{\partial}{\partial u}
 \langle h_{ij} h_{k\ell} \rangle \sum_{m = 1}^N  x_m^k \frac{\partial}{\partial x_m^\ell} \left\langle{\cal O}(x_1,\ldots,x_N)\right\rangle\,.
\end{align} 
Further details on its derivation and explicit checks are shown in Ref.~\cite{DeLuca:2024asq}.
 Equation~\eqref{CRcorrelators-sub-v2} represents the consistency relations for equal-time correlators under the subleading memory-induced symmetry in TT coordinates.

\vspace{0.1cm}
%----------------------------------------------------------------------------------------------------
\noindent{{\bf{\em Conclusions.}}}
%----------------------------------------------------------------------------------------------------
Gravitational memory effects refer to the persistent change in the GW strain caused by events like binary mergers, affecting the separation between two freely falling detectors before and after the passage of GWs. These effects are closely tied to the BMS symmetry group of asymptotically flat spacetimes, representing transitions between different BMS frames via supertranslations.

In this work, we explored the form of residual diffs in the local TT frame around a GW detector, such as LISA. In this gauge, the residual diff that 
generates a constant TT mode, describing the shift induced by memory effects, corresponds to a volume-preserving, anisotropic spatial rescaling. We have shown that this is completely equivalent to a BMS transformation, together with a time-dependent translation and spatial rescaling to restore TT gauge. Similarly, from Eqs.~\eqref{TT space diff} and \eqref{TT time diff}, the TT residual diff that induces a constant velocity kick is given by a time-dependent anisotropic spatial rescaling, a time-independent acceleration transformation familiar from the equivalence principle, and a spatially-inhomogeneous time translation. This is also equivalent to a BMS transformation with suitable compensating diffs~\cite{DeLuca:2024asq}.

Armed with the explicit TT residual diffs, we first derived the leading and tree-level subleading soft theorems for scattering amplitudes. They relate the amplitude of~$N$ hard modes with the~$N+1$ amplitude with a soft graviton, which corresponds to the zero-frequency limit of the memory term. Similarly, we derived the leading and subleading consistency relations for equal-time correlators involving scalar fields. These consistency relations are new, and are recognized as the flat-space analogue of the well-known tensor consistency relations in cosmology. These findings provide a unified perspective on gravitational memory, soft theorems, and symmetries, enhancing our understanding of their interplay in both cosmological and asymptotically flat spacetimes.

There are many directions of future research. An immediate goal is to generalize the soft theorems to include hard tensor modes. Analogously to inflationary consistency relations~\cite{Hinterbichler:2013dpa}, their form would be slightly more complicated, owing to the linear transformations required to preserve TT gauge. 
Furthermore, it would be interesting to extend our analysis to additional memory effects, such as the spin and center-of-mass memory, tails-of-memory (arising from loop corrections), and their interplay with soft theorems and consistency relations. 
Finally, we plan to perform a more in-depth investigation of observational prospects associated to  memory-induced soft theorems.

\vspace{0.1cm}
%----------------------------------------------------------------------------------------------------
\noindent{{\bf{\em Acknowledgments.}}}
%----------------------------------------------------------------------------------------------------
We thank D. Nichols for interesting discussions, and an anonymous referee for valuable comments regarding loop-order corrections to the soft theorems for scattering amplitudes.
V.DL. is supported by funds provided by the Center for Particle Cosmology at the University of Pennsylvania. 
The work of J.K. is supported in part by the DOE (HEP) Award No. DE-SC0013528. The work of S.W. is supported by APRC-CityU New Research Initiatives/Infrastructure Support from Central.

\bibliography{draft}

\end{document}